\documentclass{elsart}
\usepackage{graphicx}
\usepackage{latexsym}
\begin{document}
\begin{frontmatter}
\title{Optical production and detection of dark matter candidates}
\author[Pisa]{F. Brandi},
\author[Padova]{M. Bregant},
\author[Trieste]{G. Cantatore},
\author[Trieste]{F. Della Valle},
\author[Pisa]{S. Carusotto},
\author[Ferrara]{G. Di Domenico},
\author[Legnaro]{U. Gastaldi},
\author[Udine] {E. Milotti},
\author[Legnaro] {R. Pengo},
\author[Pisa]{E. Polacco},
\author[Trieste]{C. Rizzo}\footnote{present address: Universit\'{e} de Toulouse, France},
\author[Padova]{G. Ruoso},
\author[Trieste]{E. Zavattini},
\author[Ferrara]{G. Zavattini\thanksref{Author}}
\address[Trieste]{Universit\`{a} di Trieste, Italy and INFN, Sezione di Trieste, Italy}
\address[Pisa]{Universit\`{a} di Pisa, Italy and INFN, Sezione di Pisa, Italy}
\address[Padova]{Universit\`{a} di Padova, Italy and INFN, Sezione di Padova, Italy}
\address[Legnaro]{INFN, Laboratory Nazionali di Legnaro, Italy}
\address[Ferrara]{Universit\`{a} di Ferrara, and INFN, Sezione di Ferrara, Italy}
\address[Udine]{Universit\`{a} di Udine and INFN, Sezione di Udine, Italy}
\thanks[Author]{Corresponding Author: Guido Zavattini, 
Dipartimento di Fisica dell'Universit\`a di Ferrara, via 
Paradiso 12, I-56100 Ferrara, Italy}

\begin{abstract}
The PVLAS collaboration is at present running, at the Laboratori Nazionali
di Legnaro of I.N.F.N., a very sensitive optical ellipsometer capable of
measuring the small rotations or ellipticities which can be acquired
by a linearly polarized laser beam propagating in vacuum through a
transverse magnetic field (vacuum magnetic birefringence). The apparatus
will also be able to set new limits on mass and coupling constant of light
scalar/pseudoscalar particles coupling to two photons by both producing
and detecting the hypothetical particles. The axion, introduced to explain
parity conservation in strong interactions, is an example of this class of
particles, all of which are considered possible dark matter candidates.
The PVLAS apparatus consists of a very high finesse ($>$ 140000),
6.4 m long, Fabry-P'{e}rot cavity immersed in an intense dipolar magnetic
field ($\sim$6.5 T). A linearly polarized laser beam is frequency locked to
the cavity and analysed, using a heterodyne technique, for rotation and/or
ellipticity acquired within the magnetic field.
\end{abstract}
\end{frontmatter}
\begin{figure}[h]
	\centering
	\includegraphics[width=8.0cm]{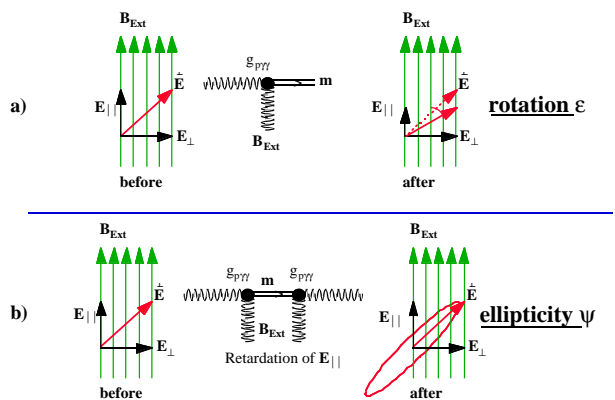}
	\caption{a) Dichroism induced by the production of a massive 
	particle coupling to two photons; b) Ellipticity induced by the 
	retardation of one of the electric field components by the virtual 
	production of a massive particle coupling to two photons.}
	\label{dich-ellip}
\end{figure}

Since any neutral light particle couples to two photons, with a strength depending
on its particular nature, a possible detection strategy can be pursued 
by optical techniques \cite{maiani}.
Among such particles there could be those considered to be possible dark matter candidates
\cite{masso}. These interactions are explored by the PVLAS experiment by sending a 
linearly polarised laser 
beam through a transverse magnetic field, and by measuring changes in 
the polarisation state of the light 
\cite{taup93,quantum,hyperfine,idm98,sandanski}.
If the electric field of the light is (parallel)perpendicular to the 
magnetic field both will couple to the (pseudo)scalar particles.
Two effects can arise: an induced dichroism and an induced 
ellipticity \cite{maiani}. For instance, figure 
\ref{dich-ellip}a 
shows the Feynman diagram for real 
pseudoscalar particle production and the resulting dichroism, while 
figure \ref{dich-ellip}b shows the retardation mechanism following from virtual 
particle production.
In the case of real production, photons polarised 
parallel to the external magnetic field will disappear leading to an 
apparent rotation of the polarisation plane. In the other case, the 
appearance and successive decay of a virtual massive particle 
causes retardation between the two components of the electric field of the laser beam.
Following ref.\cite{maiani} the acquired dichroism $\epsilon$ and 
ellipticity $\psi$ due to a pseudoscalar neutral particle
can be written
\begin{displaymath}
\scriptscriptstyle\epsilon =-\sin 
2\alpha \left(\frac{BL}{4M}\right)^{2}N\left[\frac{\sin 
\left(
\frac{L}{2}
\left(k-\sqrt{k^{2}-k_{m}^{2}}\right)\right)}{\frac{L}{2}\frac{k_{m}^{2}}{2k}}\right]^{2},\textrm{   }
\psi=\sin 
2\alpha\left(\frac{B^{2}kL}{4M^{2}k_{m}^{2}}\right)N\left[1-\frac{\sin \left(
L\left(k-\sqrt{k^{2}-k_{m}^{2}}\right)\right)}{L\frac{k_{m}^{2}}{2k}}\right]
\end{displaymath}
where $\alpha$ is the angle between the magnetic field and the light 
polarization, $k$ is the photon wave number, $k_{m}=mc/\hbar$ is the inverse 
Compton wavelength of the neutral field, $M$ is the inverse coupling 
constant, $B$ is the external magnetic induction, 
$L$ is the length of the magnetic field region and $N$ is the 
number of passages of the light across the field region. Analogous 
expressions can be obtained for the dichroism and ellipticity induced 
by a scalar field.

\begin{figure}[h]
	\centering
	\includegraphics[width=10.0cm]{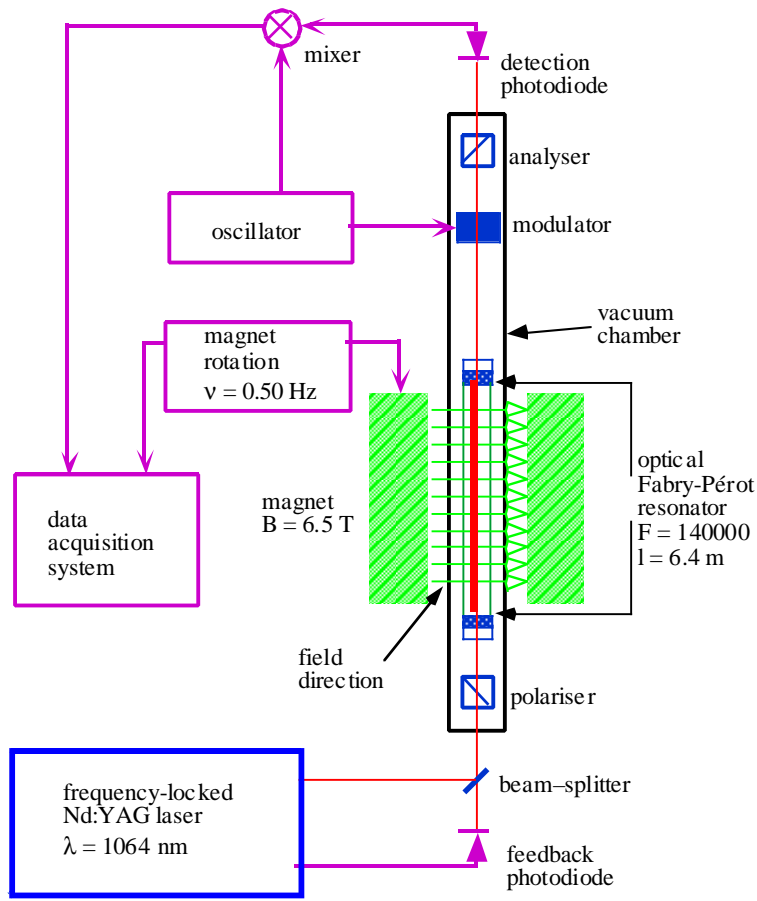}
	\caption{Scheme of the experimental setup. The linearly polarized 
	laser beam is phase-locked to a Fabry-P\'{e}rot cavity, which 
	increases the effect by $2\mathcal{F}/\pi$. The cavity passes through 
	the warm bore of a superconducting dipole magnet which rotates 
	around a vertical axis to modulate the effect. The effect is 
	detected as side bands of the modulator carrier frequency in the 
	diode signal.}
	\label{app}
\end{figure}

With the PVLAS apparatus, ellipticity and dichroism can be measured 
independently giving access to both the coupling constant and the 
mass of the produced particle. It must be noted that this detection 
method is free from assumptions on an \textit{a priori} relation between 
$m$ and $M$.
The experimental apparatus, shown 
schematically in figure \ref{app}, consists of an optical ellipsometer 
and a dipole  superconducting magnet. The ellipsometer is based 
on a vertical high finesse ($\mathcal{F}=140000$), 6.4~m long, Fabry-P\'{e}rot resonator 
cavity traversing a
1.1~m long dipole magnet which is housed in a warm bore, liquid He, cryostat 
\cite{sandanski}.
The resonator effectively accumulates photons in the magnetic region, 
thus lengthening the optical path by a factor $N=2\mathcal{F}/\pi$. The resonant 
condition is kept by means of a modified Pound-Drever-Hall frequency 
locking scheme \cite{freqlock,veryhq}.
The magnet-cryostat assembly can be rotated so 
that the magnetic field rotates in a plane normal to the light 
propagation direction, thus producing a time-modulation of the effect. 
The magnet has been energized and set in rotation 
with a field of $6.5$ T. An 
optical modulator placed after the cavity introduces a carrier frequency 
necessary for heterodyne detection.
The ellipticity and dichroism induced in the light polarisation can be 
extracted from the photodiode current by Fourier analysis.

A signal due to the existance of a (pseudo)scalar neutral particle,
which will appear in the photocurrent spectrum as sidebands separated 
from the carrier frequency by twice the magnet
rotation frequency, must necessarily be
dependent on $B^{2}$ and independent of the magnet rotation frequency. 
Also, this signal must be in the correct phase relation, depending on 
whether the particle is scalar or pseudoscalar, with the magnetic field at all rotation
frequencies. In fact the (pseudo)scalar field production is maximum for photon 
polarizations (parallel) perpendicular to the magnetic field.

The apparatus briefly discussed above is installed and functioning as 
an integrated system at the Laboratori 
Nazionali di Legnaro of I.N.F.N., near Padova, Italy. A preliminary commissioning run has 
been successfully completed in the following (not nominal) conditions: Fabry-P\'{e}rot 
cavity finesse $\mathcal{F}=100000$ (corresponding to a quality factor $Q=10^{12}$), magnetic 
field $B=4.0$ T and magnet rotation frequency of 0.43 Hz. This run has shown 
that the system is stable in time, and that data taking can be 
continuously performed as long as liquid He is available.
Preliminary data collected during commissioning yielded a sensitivity 
figure for the ellipticty of $2\times10^{-7}$ 
rad/$\sqrt{\textnormal{Hz}}$ which will allow to reach previous 
limits \cite{cameron} in a few seconds of measuring time. Taking into 
account the above experimental parameters, this sensitivity could give access to new 
physics in the yet unexplored \cite{masso,cameron} particle mass 
region $10^{-3}$~eV~$<~m~<~10^{-1}$~eV.

\end{document}